# THE APPLICATION OF THE COMPETENCY-BASED APPROACH TO ASSESS THE TRAINING AND EMPLOYMENT ADEQUACY PROBLEM

AIT HADDOUCHANE Zineb[1], BAKKALI Soumia[2], AJANA Souad[3] and GASSEMI Karim[4]

[1] Department of Mechanical Engineering, Higher National School of Electricity and Mechanics, Casablanca, Morocco
[2] Department of Electrical Engineering, Higher National School of Electricity and Mechanics, Casablanca, Morocco
[3] Department of Mechanical Engineering, Higher National School of Electricity and Mechanics, Casablanca, Morocco
[4] Department of Trade, High School of Business and Management, Casablanca, Morocco

## ABSTRACT

*This review paper fits in the context of the adequate matching of training to employment, which is one of the main challenges that universities around the world strive to meet. In higher education, the revision of curricula necessitates a return to the skills required by the labor market to train skilled labors.*

*In this research, we started with the presentation of the conceptual framework. Then we quoted different currents that discussed the problematic of the job training match from various perspectives. We proceeded to choose some studies that have attempted to remedy this problem by adopting the competency-based approach that involves the referential line. This approach has as a main characteristic the attainment of the match between training and employment. Therefore, it is a relevant solution for this problem. We scrutinized the selected studies, presenting their objectives, methodologies and results, and we provided our own analysis. Then, we focused on the Moroccan context through observations and studies already conducted. And finally, we introduced the problematic of our future project.*

## 1. INTRODUCTION

The adequacy of training to employment is essential; it is the keystone to any desired development. Graduates should be able to learn new skills, new expertise to achieve progress in a world increasingly laborious and demanding. The involvement of all parties is very important to deal with this problem and to guide young people towards the right paths [1].

The integration of the expectations of the business world into training is one of the main features of the competency-based approach. However, developing a study plan based on the skills required by the labor market remains a delicate task, because it must both take into account the expectations of the business world and implement goals that can be considered and evaluated [2].

The purpose of introducing the competency- based approach in vocational training, for example, is to ensure an efficient training while avoiding early specialization. But in parallel, we must educate graduates so as to be versatile and to excel in practice by bringing closer educational institutions and businesses. In addition, firms witness the incorporation of new technologies, which urges us to revise training programs, taking into account the requisite skills [3].

The aim of this work is to present a synthesis of the literature review discussing the adequacy of the training-job match. We begin by defining the conceptual framework that includes different concepts related to this approach. Then we present some schools that examined the issue of the training-job match in accordance with various aspects. Then we select a few works that have adopted the competency-based approach and called for its implementation; we highlight these works' objectives, methodologies and results, and we provide our analysis and perspective.

Finally, we investigate the training adequacy approach within the Moroccan context based on some already conducted studies and through the introduction of the problematic of the project we desire to carry out in the future.

## 2. CONCEPTS AND DEFINITIONS

On the one hand, we considered it essential to begin with solving the problem related terminology. This work is centered on terms that have acquired several definitions in various writings; thus such terms do not all refer to the same concepts. Therefore, we need to review the definitions to which we have referred.

We distinguish a pivotal concept, namely the adequacy of training to employment. On the other hand, we identify other peripheral concepts that affect the competency-based approach, which are tools leading to the adequacy: competence and the referential line -they constitute a framework for the competency-based approach.

### 2.1. Adequacy of training to employment

The word "adequate" means "corresponding and matching perfectly."

We distinguish between two types of adequacy: the qualitative and quantitative adequacy. As regards the qualitative adequacy, it is achieved between two elements and results in symbiosis.

Vincens [4] shows that qualitative adequacy is compared to the concept of container-contained:

"the adequate match between the container and the contained is achieved when the predefined objective is reached ... The container perfectly suits such contained in the sense that it provides the desired protection, and the contained is no less perfectly suitable for the container in the sense that it uses all the properties of the latter" [our translation] (pp. 149-150).

As to the quantitative adequacy, it is defined by the fact that each element belonging to a starter set has its complement in the target set, assuming the existence of a qualitative adequacy between the elements of the two sets [4].

The qualitative adequacy between a job and a person requires at least that such person has the ability to accomplish the tasks related to the job in order to achieve the desired result. Thus, training should be conceived as a set of clear elements adapted to the requirements of jobs [4].

Note also that we must differentiate between downgrade and inadequacy. Indeed, the term downgrade is directly related to the labor market; it is evoked when employment is below the social class or rank described, and it concerns especially beginners. While we speak of inadequacy when there are differences between the diploma and the professional field of employment. Thus, we get four ways in which the diploma matches the employment of an individual: well graded and adequate, downgraded but adequate, well graded but inadequate, downgraded and inadequate [5].

## 2.2. The Competency-Based Approach (CBA)
### 2.2.1. Competence: concepts and definition

When one refers to the English language dictionary, the word competence is explained as follows: "the ability to do something successfully or efficiently" [6], or, in French language dictionary, "a proven ability in a particular subject area as a result of the amount of knowledge possessed, and it can be assessed" [our translation] [7].

The English and French language link competence to the qualifications of a person in a given field.

Several authors have also defined the term "competence" as follows:

Aubret (as quoted in Batime 1999) defines competence as the operational 'technical skills', the knowledge (general and technical), and professional behaviors mobilized and usable in current and future work situations [8].

The definition presented by Le Boterf (as quoted in Perrenoud 2002) considers the competence as 'knowing how to mobilize' [9].

Perrenoud (as quoted in Tarek 2010) considers that the competence allows one to face a complex situation, build an appropriate response, without going into a register of preprogrammed responses [10].

According Zarifian (as quoted in Batime 1999), competence involves taking initiative and responsibility in the professional situations one faces. It is characterized by the ability to build on prior acquisitions and turn them according to different situations [8].

For Tardif (as quoted in Arena-Daigle 2006), a competence is a 'knowing-how-to-act' complex based on the effective mobilization and combination of a variety of internal and external resources within a family of situations [11].

Based on these definitions, competence can be likened to the integral mobilization of a diversity of internal resources ('knowledge', 'technical skills', and 'social/interpersonal skills') and external (material and human) to solve a given complex situation.

### 2.2.2. Competency-based approach: definition and objectives

According to Rogiers [12], the competency-based approach relies on three fundamental objectives:

- "Emphasizing the competencies that the student must master at the end of each school year and at the end of compulsory schooling, rather than stressing what the teacher must teach. The role of the latter is to organize the learning outcomes in the best way so as to bring their students to the level expected" [our translation] (p. 106).

In fact, the responsibility for learning is entrusted to the student who has to build his or her own knowledge through means made available by the teacher. The student becomes a learner who must suggest ideas first, have the desire to know and learn, organize work through using new technologies, assimilating new learning methods, and looking for new information [13].

The new role of the teacher consists in encouraging the learners to acquire the knowledge, which must be facilitated but not mechanically transmitted, and entrusting the preparation of certain tasks to the students. The teacher becomes a "facilitator" who advises the learners, motivates and encourages them to be creative, ensures the planning and organization of activities, and suggests ideas without imposing them [13].

- "It is also about giving meaning to the learning outcomes, showing the student what everything he learns at school serves. To do this, it is necessary to move beyond lists of content subjects that have to be learnt by heart ... the competency-based approach teaches them to continuously relate their learning to situations that make sense to them and to use their acquisitions in these situations" [our translation] (p. 106).

What characterizes the CBA (competency-based approach) is that teaching aims for new goals which are not related to the content to be conveyed but rather to the capacity for action achieved by the student. The latter must be able to perform a particular task by mobilizing all resources (knowledge, technical skills, behaviors) [14].

- "Finally, it is a matter of verifying and validating the student's achievements in terms of resolving concrete situations, not in terms of the sum of knowledge and know-how that the learner often hastens to forget, and which he does not know how to use in real life" (p. 106).

This means that the student will be evaluated based on their ability to act instead of being assessed in terms of the knowledge that does not affect the real situations they face.

Therefore, the CBA came into being as a result of two needs. First, the corporate world wanted a workforce having adequate training, and second, there was a need for pedagogical concepts centered on the individual result instead of abstract knowledge [14].

We had to combine two concepts: on the one hand, the focus on the skills development and, on the other hand, the effective use of compulsory education in favor of economy [15].

Certainly, the CBA has initially known significant prosperity on the theoretical level, without being affected by the corporate world. However, it has been proven afterwards that such an approach is of great importance to business circles, which justifies most likely its current success [14].

The impact of CBA is outstanding in the domain of training and education. Indeed, in most Western countries, curricula are formulated based on the requisite skills [13]. According to Perrenoud (1997), a competency-based approach needs a complete reconstruction of training systems and processes [16].

### 2.2.3. Referential line

The referential line is considered as a framework that underpins all elements of training in the CBA, starting from the skills required by the labor market unto the way they should be assessed. This line includes four frames of reference presented in Figure 1.

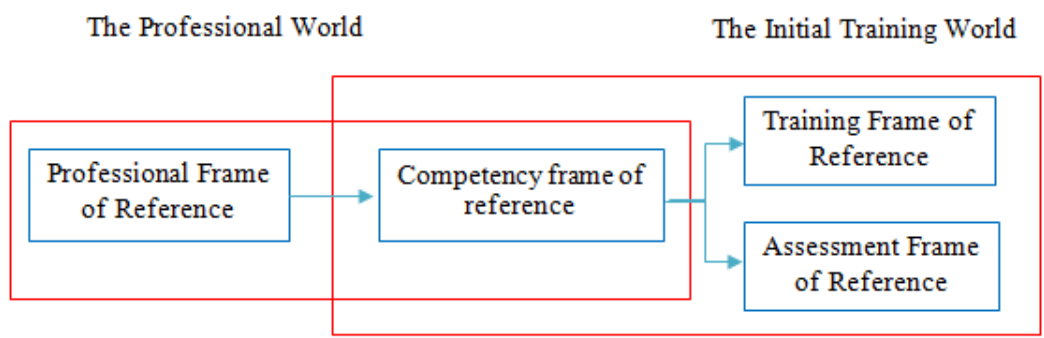

Figure 1. Logical organization of the various standards used in initial training [Our translation] (p. 114) [17]

- Professional Frame of Reference

The professional frame of reference contributes to determining the essential skills to perform the assigned functions, as well as the competences to be developed after the emergence of unprecedented activities [18].

The creation of a professional frame of reference is considered a vital phase, especially when identifying the skills and the conception of the training and evaluation systems to implement. It is a mediating tool between the various parties, since the trainers involved in its conception are able to determine the professional purpose of training, while professionals can define actual situations and requirements that are necessary to practice the profession in the future. A professional frame of reference mainly contains descriptions of the functions[1], activities[2] and tasks[3] related to the practice of a profession [17].

- The Competency frame of reference

The competency frame of reference constitutes "the basis of the pedagogical part of work, in other words, it is the interface that makes it possible to move from the world of work to the world of training" [our translation] (p. 114) [17].

It associates the requirements of the profession with the individuals from the perspective of human resource management, of both training and evaluation.

---

[1] A function allows to justify the presence of a certain activity and what it contributes to. Its determination is important because it provides the basis for identifying key activities [17].

[2] Key activities help define the way in which each mission should be accomplished. It corresponds to the overall level of the description of a job [17].

[3] The tasks relate to the operations that should be performed In order to carry out an activity. They describe what an individual should do but giving far more detail the activities [17].

A competency frame of reference is then constituted of a list of essential skills that must be assimilated to perform the tasks of a given job. According to the Pedagogy of Integration, it must also develop skills for future use, both in training and in the validation of acquisitions.

To make the transition from the workplace to the world of education, we need the essential and evaluable skills that provide a clear view of the key dimensions of the profession. It is important to evaluate through complex professional situations instead of mere knowledge and technical skills, in order to ensure the relevance of the evaluation [17].

The competency frames of reference emphasize individuals, their cognitive skills, and their individual abilities through imitable statements. Thus, they are set up as canonical writings serving to create harmony between practices and representations [19].

- The Training frame of reference

The curriculum or Training frame of reference, according to Demeuse and Strauven (2006), is a document that provides clear vision of pedagogical guidelines in an organized, directed and chronological way, and through which learning will be managed in relation to goals. It includes several elements, such as the subject-specific contents, the learning outcomes to be set up, the teaching and learning processes, the goals to be achieved and the evaluation features,… [20].

## 3. ADEQUACY OF TRAINING TO EMPLOYMENT
### 3.1. Different schools having treated the training-job match

Several studies have discussed the approach of the adequacy of training to employment from different perspectives and using various methodologies.

Champy-Remoussenard [21] highlighted the characteristics of the links between education, training, and employment, taking into consideration the economic development. She focused on three strategies to examine the reorganization of the relationships between the aforementioned three concepts, namely: the validation of experiential learning, the alternation and the development of internships, and the entrepreneurial model of activity and training.

An article entitled "When training is not enough: preparing students for employment in England, France and Sweden" brought into light another aspect of the link between jobs and rarely deepened studies. Charles [22] attempts to prove, through his research, that higher education in England and Sweden does not really opt for the concept of professionalizing students, unlike what most believe. He also asserts that promoting the assimilation of professional skills within the training process distinguishes the French model, and that the multiplication of "professionalizing" internships and trainings is a new way of interpreting the concept of the "adequacy of training to employment" that particularly characterizes France.

Through their work "pedagogical tools in higher education and graduates' employment", Chevallier and Giret [23] showed interest in the impact of the pedagogical systems employed in teaching on the start of the professional careers of graduates. They proceeded by conducting a survey among young Europeans who recently graduated so as to classify the types of the pedagogical tools with reference to the results of the questionnaire used.

Two successive articles fit in the context of the competency-based approach practices in order to meet the expectations of the business world. Deschryver et al., [2] and Deschryver, Charlier and Fürbringer [24] intend to achieve adequacy between the training and the employment of engineers through the development of a competency frame of reference. In these articles, they present the results of a practical application of the competency- based approach in the mechanical engineering section of the Ecole Polytechnic Federal School of Lausanne (EPFL).

Another research addresses the issue of competences required for the entrepreneur by conducting a field study so as to develop a frame of reference. In their article, Loué and Baronet [25] aim to validate a range of competences required for the entrepreneur.

A study by Santelmann [26] intends to elucidate the differences between the desired goals to be achieved through training and the skills produced during the practice of the profession, using as example the profile of technicians and supervisors of the industry that is experiencing a polymorphism.

Felouzis [27] advocates the proposition that links the quality of degrees and diplomas with the professional integration in the labor market. He highlights the uncertain character that defines the labor markets and higher education.

In this work, we will scrutinize studies that have tackled the issue of the adequacy of training to employment, opting for the solution of applying the competency-based approach. In addition, we will investigate a study that compared the training systems in diverse countries. Indeed, the competency-based approach is characterized by the versatility of the teaching methods on the one hand, and on the other hand, by a wide array of activities that replace the traditional lecture-based courses, such as case studies, scenario-based teaching ... This approach is now applied in many corners of the world (US, Australia, Europe, ... etc.) and has become "the brand of the new educational policies supported by the UNESCO, the OECD and the States involved in the Bologna Process, that is aiming to render the dissemination of knowledge an engine for economic and social development" [our translation] (p. 16) [28].

It is an approach known for its variety of uses in the field of training design and the evaluation of training programs. It highlights, in the best way, the appeal to knowledge, the understanding of competencies, more consistent evaluation methods, and re-acquisition of the training course by the individual [28]. In addition, it has a harmonious structure of curriculum components, such as the modes of the learning assessment, the textbooks, the training programs ..., which guarantees efficiency in the curricula. It also provides a holistic education because it does not focus only on the cognitive aspects but is interested in major societal challenges of nowadays. This perspective seeks to involve all parties to be responsible, including students, and aims to create relevance in the curricula. The competency-based approach also provides inclusive education since it does not exclude any type of student from the schooling process, thus ensuring equity [29].

In order to apply the competency-based approach, surveys were conducted, along with the companies, to determine the acquisition levels or the in-depth study of a competence. When the latter is developed through learning, Bloom's taxonomy proves to be of great importance to the identification of these levels.

The objectives of the cognitive domain were classified by Bloom's taxonomy on the basis of six hierarchical levels, going from the simplest to the most complex one [30], [31].

Figure 2 presents the pyramid that illustrates these levels.

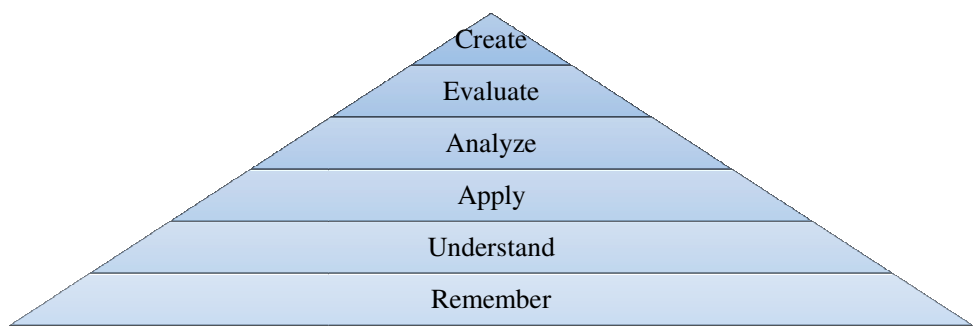

Figure 2. Bloom's Taxonomy [31].

- « Remember » is the first level that involves gathering information and handling it in a very basic way.
- « Understand » is the step that consists in examining the information by giving an example or by rephrasing it.
- « Apply » is the third level in which knowledge and strategies are mobilized into classical situations.
- « Analyze » is the step that focuses on the understanding of components and of the functioning of the tool or the method.
- « Evaluate » consists in improving the existing methods or in creating new ones, when the ordinary rules do not work properly.
- « Create » is the last stage which entails forming hypotheses, estimating results, explaining a decision, and debating [30], [31].

Every teacher must incorporate a taxonomy at the level of learning, assessment and teaching. Indeed, Bloom's taxonomy is used by educators and curriculum developers. Its relevant structure makes it an effective tool for planning and taking into account the different levels of thinking within courses, teaching units and training programs. Bloom's taxonomy has allowed for the creation of a common language among teachers [32]. Moreover, it is known for its usability in various situations, which favors its application in the modeling of the learning goals in education [33].

## 3.2. Synthesis of the literature review discussing the 'training-job match' approach

In this part, we will introduce the objective, the methodology and the results of some works that tackled the problematic of the adequacy of training to employment.

The comparative study by Charles [22] was conducted as an experiment among students in two schools in different countries: a prestigious management and economy school and a democratization school that offers history courses.

The author started by determining the mechanisms to which the students are subject to during the passage from studies to employment in England and Sweden. Then, he shed light on the professional aptitude of The French students. Finally, he clarified the particularity of the French model by pinpointing the basics of the "adequationist" thinking [22].

The results were presented, for each country separately, as follows:

In England, students' "employability" is the purpose of higher education. The objective of this notion is to build a set of cross-disciplinary skills that will enable students to be employment material in the majority of management jobs, instead of developing skills that suit one specific professional job during the training, since this notion is of no significance vis-à-vis students and employers. Thus, education in England does not allow having a specific professional identity [22].

In Sweden, higher education is based on the interweaving of training and employment. Indeed, there are two forms of professional experiences: either students alternate periods of study and periods of work, or they work part-time alongside their studies. The purpose of this interweaving is to push students to think about their professional future and to have the necessary professional experiences during their university course. This will allow them to integrate jobs that are closely related to their studies.

The professional qualification originates from the training because the knowledge and the theoretical background acquired at university develop all the more in a work environment. In this way, the Swedish model persistently adopts the concept of professional qualification that promotes learning general knowledge without overlooking the acquisition of other specific skills [22].

In France, higher education proceeds through three stages: studies, professional integration and access to the labor market. Unlike England and Sweden, France puts more emphasis on professional qualification [22].

French students have two methods of preparation for employment, which consist in two types of training; on the one hand, professionally-oriented training which implies internships, and on the other hand, civil service contests and school professional training.

In fact, internships are highly undergone in France –roughly 74% of graduates have completed internships as part of their university program. These are considered mandatory for the validation of degrees in the majority of trainings. Moreover, they can take up to a year, and their objective is to allow the student to have a job and take the same responsibilities as an actual employee [22].

The concept of the adequacy of training to employment is more and more discussed, given the risk of the unemployment rate growth. Its rationale of basing on individuals capacities seeks to guarantee a good match between academic degrees and jobs.

Nowadays, the majority of graduates integrate the private sector. Such change in terms of beneficiary has affected the system of higher education and forced universities to change their training content [22].

The English model gives no importance to the curricular contents emphasizing professional skills that are required by the employers. It is unlike the French model that usually tends to meet the expectations of the labor market. Thus, the French model illustrates more the adaptation of trainings to jobs. However, in England, the concept of "employability" is taking another direction towards the professionalization of higher education, since professional integration has now become more and more difficult [22].

The project carried out by Deschryver et al., [2] consists in elaborating study plans with respect to competencies. It has been realized thanks to the support of the Rectors Conference of the Swiss Universities and in collaboration with teachers. This project is meant to deal with all issues relating to the engineer's profession, notably the cognitive, interpersonal skills, and technical skill. It will also guarantee coherence between the objectives of the training, the courses, and their evaluations. Moreover, it is supposed to help students take the right direction and gather teachers together around a project whose main interest is to train competent mechanical engineers.

In the EPFL, a questionnaire has been designed to investigate 37 different corporations in an attempt to have a clear vision of the competencies required in the labor market. To complete this task, they resorted to the model of competencies referred to by Le Boterf (2006) and Tardif (1999). The resources framework contains three axes, namely, the thematic axis dealing with knowledge and technical skills; the thematic axis dealing with interpersonal skills and the other axis, inspired by Bloom's three-level model, is the deepening: (1) remembering and understanding, (2) applying and analyzing, (3) creating and evaluating. The questionnaire is then interested in three categories of responses, which are: the interest of a particular profession, the desired level of depth, and a justification for the answer [2].

The quantitative analysis of such questionnaire has shown that fundamental sciences should be acquired to reach application level, engineering sciences to achieve optimal mastery, and human and social sciences to attain knowing-understanding level. Hence, the people who responded to the questionnaire showed a major interest in engineering science and particularly in three themes, namely, design, knowledge of the materials, and thermodynamics. The analysis has also allowed for a classification of the different technical skills that should be considered essential by the students, according to the order of demands by the labor market. As regards the interpersonal skills, there are corporations that consider them sine qua non for the integration of an engineer into the professional domain; others believe that by making efforts, such skills can be acquired since the first job [2].

Concerning the qualitative analysis of the results, they arranged the given comments into categories that were agreed upon after the first reading, for example, knowledge, the reality of the profession, etc. Then, they were rearranged in accordance with the competency dictionary of civil engineers of the Université Libre de Bruxelles so as to make use of a research that was conducted earlier. Consequently, the result of this qualitative analysis is a list of competencies along with their components illustrated by extracts of the enquiry [2].

Mechanical engineering is divided into different subject areas; therefore, speaking common language was a real challenge. Consequently, three types of documents were developed so as to direct the study plan: the first identifies the competencies required for an EPFL mechanical engineer, their components and the links between them; the second concerns the way in which the competencies expected of the training would be dealt with within each subject area; while the third relates to the enhancement of the required competencies during the training period [24].

Then, they concentrated on three main parts so as to ensure harmony within the curriculum:

On one hand, the links between the competencies and the outcomes of learning. In fact, a series of competencies and their components was elaborated in collaboration with teachers on the basis of the enquiry and an existing competency dictionary. These competences are interrelated and have a limited number of components ranging from three to five, and can also be applied to other engineering programs. So as to determine the outcomes of certain competences, they had recourse to a Canadian competency dictionary [24].

On the other hand, they have focused on the articulation between subject areas and professional profiles. In fact, two kinds of development that were achieved are, first, the specification of the subject areas and, second, the provision of a descriptive account relating to typical business situations. To attain such goal, a team of teachers have worked out a project in which they defined their domain, such as the key concepts, the domains of application, etc. Besides, each team has determined professional situations relating to its domain as well as the required competencies that must be achieved with respect to each situation [24].

Finally, there was the part of the links between learning outcomes and learning activities, where teachers decided to connect each learning outcome with a suitable learning activity or situation. In order to develop the project, the teachers have also considered the suggestion of new activities or situations for the learning outcomes which have not been originally dealt with [24].

With regard to the study conducted by Loué and Baronet [25], which aims at preparing an entrepreneur's dictionary of competencies, the adopted methodology was initiated by a review of literature on the subject, so as to suggest a preliminary set of theoretical competencies for the entrepreneur. Then, a first list of competencies was presented to a sample of entrepreneurs for approval, and that is through a set of qualitative interviews and subsequently via a quantitative questioning in order to end with a significant sample.

In the first stage, the focus was on the entrepreneur competencies according to various authors so as to ensure a proper division of the competencies axes. Then, concerning the qualitative validation, interviews were conducted with 29 business executives working in different sectors. This phase was divided into two parts: the first was brainstorming-like so as to collect a maximum of competences to be mastered; the second was based on a well-prepared questioning that was provided to the interviewee; such questioning comprised all determined axes of competencies [25].

In the second study, the interviews analysis allowed the validation of 70% of theoretical competencies proposed earlier in the dictionary, and more than 43% of additional competencies have come up in the totality of axes. Ultimately, a list of 55 competencies was obtained to constitute the foundation for the quantitative phase. The latter was completed in three countries, France, Algeria and Canada (Quebec) with 402 entrepreneurs. The formalized competencies were transformed into a questionnaire. And after having conducted a statistical analysis, a total of 44 competencies have been retained [25].

This type of research did not only result in creating an operational dictionary of competencies but also in validating it inside French-speaking countries, in an endeavor to ensure a great deal of continuity. Moreover, the outcomes of such research represent an input for pedagogical engineering, as well as a verification of the adequacy of training curricula with regard to the demands of the labor market [25].

### 3.3. Summary table

Table 1 presents a recapitulation of the synthesis about the state of the art of the training-job relevance approach, especially, studies which we chose to detail.

Table 1. Summary table of some studies

| Author | Year | Objective | Methodology | Results |
|---|---|---|---|---|
| Nicolas Charles | 2014 | - To focus on a new aspect of making trainings adequate to employment. | - To compare the training systems of England, Sweden, and France. | - The French model is the one that takes into account the satisfaction of the labor market needs. On the other side of the coin, the English model cares only about cross-curricular competencies without showing interest in the curricula that shape professional competencies.<br>- The Sweden model advocates the idea that acquired knowledge becomes more effective in the professional field and hence opts for the interweaving of trainings with jobs. |
| Nathalie Deschryver, Bernadette Charlier, Jean-Marie Fürbringer and Remy Glardon | 2010 | - To produce a frame of reference for Bachelor and Master programs based on the competencies expected of an EPFL mechanical engineer. | - To investigate 37 corporations via a questionnaire.<br>- To analyze the results in a qualitative and a quantitative way.<br>- To establish the links between competencies and learning outcomes, the articulation between subject areas and professional profiles, and connecting learning outcomes to learning activities. | - A well-detailed dictionary of competencies (macro-competencies and their components, description, situational, training and evaluation instances, and typical professional situations...) and a new curriculum for Mechanical engineers of EPFL. |
| Nathalie Deschryver, Bernadette Charlier and Jean-Marie Fürbringer | 2011 | - To attain a better training-job match for the engineer. | | |
| Christophe Loué and Jacques Baronet | 2011 | - To prepare a dictionary of competencies for entrepreneurs. | - To draw up a list including the entrepreneur's theoretical competencies.<br>- To validate this list through questioning a bunch of entrepreneurs via an interview as well as a questionnaire for the sake of a more significant sample. | - An entrepreneur's competencies dictionary that is operational and validated in three French-speaking countries (Algeria, Canada-Quebec and France) serving as a foundation for the design of new training programs. |

The elements of this synthesis related to the practices of the competency-based approach and the adequacy of trainings with regard to jobs intertwine with the notion of frames of reference, which demonstrates the inseparable nature of its different concepts. The two last studies similarly aim at the enhancement of the training components, that is, by referring to a competency frame of reference established after investigating different companies. The outcome of this project carried out by the EPFL was presented in the form of a competency dictionary for mechanical engineering. It is characterized by its simplicity, which makes it intelligible even to laymen. Besides, it has a generic and thus transposable aspect, but also specific features dealt with in terms of the outcomes of learning. Another asset is the framework of the study plan founded upon simple axes that are similar to Bloom's taxonomy. A detailed account of the competency dictionary's components and the links between them was provided in the document.

The main characteristic of the entrepreneur's competency frame of reference is that business founders ratified it in three French-speaking countries, which widens its operational scope. However, this frame of reference only contains macro-competencies and their components without further clarifying the instances of professional and training situations.

The adequacy of training to employment is a recurrent problematic in various countries around the world, including Morocco. The latter has witnessed an unprecedented economic growth during the last decade. In order to keep up with such progress, it is necessary to reassess and update training programs and increase the number of majors in terms of initial and vocational education.

## 4. THE MOROCCAN CASE

Moroccan universities provide training for 90% of students, which urges them to guarantee high-level employability trainings to its laureates. The adequacy of training to employment is of great importance in order to achieve efficiency in the educational policies. The quality of educational systems and training proves significantly useful notably after the great impact that globalization has had on the labor market. As a result, the improvement of such quality requires defining the competencies expected by companies, comparing them with the instructed competencies, and ultimately suggesting solutions for reducing the discrepancies [34].

Our problematic issue starts from the following findings quoted in the frame 1:

- In 2012, Lachcen Daoudi, Minister of Higher Education, Scientific Research and Staff Training, stated **"Morocco suffers from a significant lack of skilled labor force. 7000 engineers are immediately needed to support the major projects launched."** [our translation] [35].

- As of May 14, 2015, the Party of Justice and Development wrote on its website that during a conference held in the National School of Applied Sciences of Tetouan, the same minister highlighted the fact that **trainings** meant for **engineering students** should be **relevant** to the **needs of the labor market,** keeping an eye on the ongoing **industrial development** as well as the **economic changes**.

- In 2016, Lahcen Daoudi announced that Morocco **"always manages the quantitative aspect; our challenge is qualitative one. Our trainings are not yet satisfactory"** [our translation] [36].

Frame 1. Findings about the adequacy of trainings to employment

Based on these findings, we deduce that the new challenge of Moroccan universities, especially Higher Schools of Engineering, is to have a qualitative adequacy of trainings to business profiles. Competent human resources will considerably contribute to the attraction of investors, to progress, and to business competitiveness.

There are some national works that have already taken interest in the subject of adequacy of trainings to employment.

A report written by Ouahab [37], an expert in engineering training, showed that the Moroccan unemployment rate increases with the increase of the educational level (74,4% of higher education graduates are long-term jobless), which confirms the inadequacy of trainings with regard to employment and the ineffective increase of qualified jobs. He confirms that the irrelevance of trainings to the competencies that employers require, the poor quality of training, the lack of knowledge of the labor market, and slow economic growth are some of the difficulties that young people confront to access the labor market... Afterwards, he laid emphasis on a number of solutions which must be implemented, such as keeping track of the development of the labor market: which is necessary for the good management of educational and training systems, involving professionals: which will allow for the evaluation of the educational programs and the teaching methods and contribute to their enhancement... [37].

There is also the article by El Mendili [38] that considers companies to be a customer of training. It also affirms that the institution offering the training should involve the administrative staff in the methodology selection since they are the ones to specify the objectives matching with the customer's policy and identify indicators for measuring and optimizing customer satisfaction. Then, the training institution must define its potential customers and integrate them by identifying their expectations in terms of competencies and transforming them into requirements to be taken into account during the process of enhancing the training content. Besides, the administration may refer to other methods, such as engaging professionals in the training by hosting occasional lectures, seminars…, arranging periodic meetings with companies in thesis defense days…. Therefore, the training institute should revise its training curricula on a permanent and organized basis by using quality improvement tools such as Quality Function Deployment (QFP). The study of the customer's feedback as well as the economic development should not be disregarded during the update of training contents [38].

The objective of another study conducted by Saadi, Fakhraddine and Adil [34] was to improve the quality of electrical engineering training of a Moroccan school of engineering through the QFD approach.

This study focused on companies operating in the sector of electricity/electronics of National Federation of Electricity and Electronics (FENELEC). The enquiry targeted 30 organizations, that is, 10 large companies and 20 small and medium enterprises (SME) after they had been randomly selected. The first stage consisted in brainstorming sessions with FENELEC representatives to gather ideas about the actions which must be taken in order to prepare students for succeeding in their first job, deepening their knowledge vis-à-vis the company, improving the contents and objectives of the training and increasing the graduates' employability. The second stage was the identification of the various competencies desired by corporate customers through an enquiry conducted in the top-three companies. Subsequently, these competencies were divided into three classes: scientific and technical competencies, managerial competencies, and cognitive competencies. The final questionnaire was the result of former stages, and it included 48 questions divided into three groupings. The results show that companies attribute a value of 23% to managerial competencies with more emphasis on environmental and security management, a value of 43% to cognitive competencies highlighting dynamism and discipline, and a value of 34% to technical and scientific competencies with increased attention to computer science [34].

The other phase of the study consists in the application of QFD method. In fact, this method can express the customer's needs in the designer's language, in order to make a product or provide a service without additional costs. To apply QFD method to this study, the marketplace needs in terms of competencies would present the WHATs of QFD matrix and the disciplines specified in collaboration with the costumers would form the HOWs. And as they selected three categories, three QFD matrices were obtained. Furthermore, it is essential to quantify the relationship between each WHAT and HOW by a weight 0, 1, 3 or 9, depending on the goal of such relationship. The outcome of such quantity by relative weights of WHATs allows for obtaining the weights of HOWs. This will make deciding on the importance of each subject easier when programming the training design. The degree of importance characterizing each subject will manifest in the training hours as well as the coefficients of exams [34].

Before thinking about making improvements on training curricula, it is necessary to define discrepancies among curricula and the needs of the marketplace. Engineers, a vital part of human resources, contribute to the development of companies. This is apparent from the multiple and diverse tasks and responsibility which they are usually assigned.

We will begin our next research work on the adequacy of engineers' training to employment, taking as a sample Mechanical Engineering in the Higher National School of Electricity and Mechanics (ENSEM), by answering the three following questions:

- Does Mechanical Engineering training in ENSEM meet the needs of the labor market?
- What are the competencies that characterize the mechanical engineer of ENSEM? And what competencies should be developed?
- What action plans could be suggested so as to adequately fit professional profiles and deal with globalization?

The objective of the project that we endeavor to complete is the improvement of the training quality for engineers in mechanical engineering in ENSEM.

Therefore, we decided to adopt, firstly, an inductive approach. We began by conducting an exploratory qualitative study, comparing some professional profiles to the training programs of such engineering. We found that this training program meets about 90% of the needs of the labor market, which allowed us to formulate our hypothesis [39]. Then, we turned to use a hypothetico-deductive approach to either confirm or reject our assumption by means of a quantitative study. We thus started such study by comparing some international Competency Frameworks of Mechanical Engineering based on criteria identified after a bibliographical study was carried out. We used the compatibility matrix method to choose a competency framework that will form the basis of our questionnaire [40].

The latter will be used to conduct an enquiry among superiors of ENSEM graduates in 2013, 2014, and 2015. Its aim is to identify the degree of importance and the level of proficiency of each competency based on a well-defined scale.

The various results of such questionnaire will go through a statistical study in order to find out about the discrepancies between the training and the marketplace and consequently to confirm or reject our assumption. We will also develop plans of necessary actions in order to adequately match the different professional profiles and create a competency dictionary for ENSEM.

## 5. CONCLUSION

Throughout this article of synthesis, we were able to identify concepts that constitute the basis for the various works interested in the adequacy of training to employment. We also brought to light several currents that dealt with this notion. Afterward, we selected three studies and defined their objectives, methodologies and results. On the one hand, our review of literature synthesis has probed into the practical side of the competency-based approach, indicating various methodologies for establishing competency dictionaries. On the other hand, we have discovered another approach related to the adequacy of training to employment, which was the outcome of a comparative study of the training systems of three French-speaking countries. Then, we have contextualized the training-job match within the Moroccan setting through already completed works. And, finally, we have introduced the study we will work on in the future.

The next step will be a questionnaire-based enquiry conducted among Moroccan manufacturers that have recruited mechanical engineers from the Higher National School of Electricity and Mechanics (ENSEM). The goal is to reveal the discrepancies between the competencies required by the labor market and the competencies acquired by such engineers. Consequently, we will bring improvements to the training curricula of mechanical engineering in ENSEM.

## Authors

### Zineb AIT HADDOUCHANE

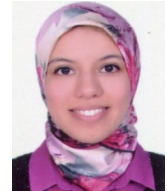

- PhD Student, Research team on Engineering Education at ENSEM, Engineering Research Laboratory (LRI), Hassan II University Casablanca, Morocco.
- Industrial and Logistic Engineer in 2014, National School of Applied Sciences

### Soumia BAKKALI

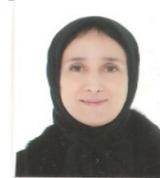

- PhD in micro-electronic from the HassanII University Casablanca in 1996.
- Since 1997, professor at ENSEM.
- Member of research team on Engineering Education and member of the system architecture team at ENSEM, Hassan II University, Casablanca, Morocco.
- She gives the following courses: analog electronic, fiber optic telecommunication and physic of semiconductor.

### Souad AJANA

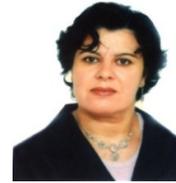

- Since 1987, professor at ENSEM, Hassan II University Casablanca, Morocco.
- Pedagogical Skills in Industrial Hydraulics, Control and Measurement Techniques and Materials Resistance
- Head of the Laboratory of Metrology at ENSEM
- Fields of research:
  Hybrid bearings in laminar and turbulent regimes
  Ex Member of the Rheology and Plastic Materials Research team
  Since 2014, Creation and Head of the research team on engineering education (ERFSI) attached to the Engineering Research Laboratory (LRI)
- In 1987, PhD in Fluid Mechanics, Lille 1 University

### Karim GASSEMI

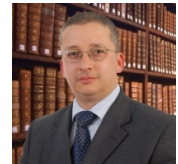

- April 2016: Laboratory Head – LAMSO (Laboratoire d'Analyses Marketing et Stratégiques des Organisations), High School of Business and Management, Hassan II University Casablanca, Morocco
- Since January 2008, Professor at High School of Business and Management, Hassan II University Casablanca, Morocco
- In 2007, PhD in Management Science, Pantheon Assas University
- Since 2003, Consultant
- In 1996, MBA, BBA (business management) – MBA (Information System), Laval University